\documentclass[12pt]{iopart}
\usepackage{amsfonts}
\usepackage{graphicx}
\begin{document}

\title[]{Exploring quantum dynamics in an open many-body system:
Transition to superradiance}
\author{Alexander Volya\dag\ and Vladimir Zelevinsky\ddag}
\address{\dag Physics Division, Argonne National Laboratory, Argonne,
Illinois 60439}
\address{\ddag National Superconducting Cyclotron Laboratory and
Department of Physics and Astronomy, Michigan State
University, East Lansing, Michigan 48824}
\begin{abstract}
We study the dynamics of a complex open quantum many-body system.
The coupling to external degrees of freedom can be viewed as a
coupling to a radiation field, to continuum states or to a
measuring apparatus. This perturbation is treated in terms  of an
effective non-Hermitian Hamiltonian. The influence of such
coupling on the properties of the many-body dynamics is discussed,
with emphasis on new effects related to dynamical segregation of
fast and slow decays and the phase transition to Dicke
superradiance. Relations to quantum optics, continuum shell model,
theory of measurement, quantum chaos, percolation theory, and to
quantum reactions are stressed.
\end{abstract}
\maketitle

\section{Introduction}

The interplay of complex intrinsic dynamics with radiation,
coupling to the continuum or to any other external influence, such
as measurement, is an important subject for quantum optics,
nanoscale devices,  theory of decoherence and quantum complexity,
future quantum computing, and physics of nuclei far from
stability, to name just few branches of modern physics, seemingly
far remote from each other. After all, the complete isolation of a
``system'' and the ``external world'' is never possible. The
chaotic and regular aspects of many-body physics, as well as
openness and decay properties, are deeply rooted in valuable parts
of Wigner's legacy. In this paper, commemorating Wigner's
contributions, we would like to stop at the crossroad of ideas related to
the complexity of many-body physics and the
Weisskopf-Wigner damping theory. We consider a mesoscopic
many-body system with internal dynamics governed by the mean field
and residual two-body interactions. We assume that the system is
perturbed by non-Hermitian terms in the Hamiltonian. These terms,
in general, model an effective coupling to the external degrees of
freedom such as the continuum of scattering states, radiation
fields or measurements.

The subject of this work is cross-disciplinary and the approach
implemented here is widely used in atomic, molecular and nuclear
physics, see our recent work \cite{continuum} and references
therein. For a broader audience we structure the paper and conduct
our discussion in a pedagogical manner. We start our work by
presenting in Sec.~\ref{sec:heff} a well known description of the
internal dynamics in an open many-body quantum system with the aid
of the non-Hermitian and energy-dependent Hamiltonian. A simple
example of two interacting spins in the presence of decay is
discussed in Sec.~\ref{sec:spins}. Although this example has no
complexity associated with the many-body dynamics, the important
physical features related to decay, as well as mathematical
aspects, are already there. This case serves as an introduction to
our main discussion presented in Sec.~\ref{sec:big} and based on
the consideration of a system of interacting fermions coupled to a
decay channel. Concluding remarks are given in Sec.~\ref{sec:conclusions}.

\section{Effective Hamiltonian of an open system}
\label{sec:heff}

The technique using a non-Hermitian and energy-dependent effective
Hamiltonian is rather common and dates back to the
Weisskopf-Wigner damping theory \cite{weisskopf30}, works of Rice
\cite{rice33} and Fano \cite{fano35}, and Feshbach projection
formalism \cite{feshbach58}. It is assumed that the many-body
states in the system can be separated into two classes, the
internal (intrinsic) states and external ones. The set of internal
states $|\alpha\rangle $ is generally understood to be composed of
the states where all particles occupy bound mean-field orbitals.
The fact that the system is open requires one to include into
consideration a set of external many-body states. These states
$|c; E\rangle $ are viewed as decay or reaction states and contain
particles in the continuum or radiation quanta; we label them with
a continuous variable of energy and a set of other quantum numbers
characteristic for a channel $c$. Discussion of the reactions is
not our primary goal in this work, nevertheless we should
emphasize Wigner's significant contribution to this subject
\cite{lane55,teichmann52,wigner48}.

Although there exist many different formulations of this
technique, here we start from the Schr\"odinger equation
\begin{equation}
H|\Psi\rangle =  E |\Psi \rangle,               \label{1}
\end{equation}
where the full wave function is a superposition of internal and
external states,
\begin{equation}
|\Psi \rangle = \sum_\alpha C_\alpha |\alpha \rangle + \sum_c \int
dE \,\chi^c_E|c; E \rangle.                   \label{2}
\end{equation}
Eliminating the external states we come to the equation for the
coefficients $C_\beta$ that describe the internal part of the wave
function,
\begin{equation}
\sum_\beta \left [\langle \alpha|H|\beta \rangle + \sum_c \int dE'
\frac{\langle \alpha|H|c;E'\rangle \langle c;E'|H|\beta\rangle
}{E-E'} -\delta_{\alpha \beta} {E} \right ] C_\beta =0.
                                              \label{incoef}
\end{equation}
This equation looks like a traditional eigenvalue problem with the
effective Hamiltonian matrix in the intrinsic space defined as
\begin{equation}
\langle \alpha |{\cal H}({ E})|\beta \rangle = \langle
\alpha|H|\beta \rangle + \sum_c \int dE' \frac{A_\alpha^c(E')
{A_\beta^c(E')}^*}{E-E'},                         \label{heff1}
\end{equation}
where we introduced $A_\alpha^c(E)=\langle \alpha|H|c;E\rangle\,,$
the coupling amplitudes between internal and external states.

The first term in Eq. (\ref{heff1}) is the standard Hamiltonian
matrix that would describe our system if it were decoupled from
the external world. For definiteness, we assume that this part of
the Hamiltonian is
\begin{equation}
H=\sum_1 \epsilon_1 a^\dagger_1 a_1 + \frac{1}{4}\sum_{1 2 3 4}
V_{1 2; 3 4} \, a^\dagger_1 a^\dagger_2 a_3 a_4 \,,
                                          \label{hamiltonian}
\end{equation}
with single-particle energies $\epsilon_1$ and residual two-body
interactions with antisymmetrized matrix elements $V_{1 2; 3
4}\,.$ The indexes ${\sl 1},\dots,$ contain a complete set of
single-particle quantum numbers, and the total many-body basis
state $\alpha$ can be constructed, in the case of
Fermi-statistics, as a Slater determinant of these single-particle
states.

\begin{figure}
\begin{center}
\includegraphics[width=10 cm]{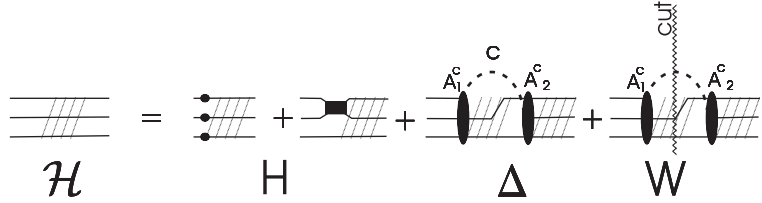}
\end{center}
\caption{\label{scheme1}A schematic representation of the
effective Hamiltonian of an open system. The Hermitian shell model
part $H$ is shown by the two terms following Eq.
(\ref{hamiltonian}), single-particle energies and two-body
residual interactions. The term $\Delta$ includes all channels $c$
while the term $(-i/2)W$ is restricted to the channels open at a
given energy.}
\end{figure}

The integral in Eq. (\ref{heff1}) can be further decomposed into
its Hermitian part (principal value) and the remaining
non-Hermitian part,
$$
\sum_c \int dE'\,\frac{A^c_\alpha(E') {A^c_\beta(E')}^*}{E-E'}=
$$
\begin{equation}
\sum_{c} \,{\sl P} \int dE' \,\frac{A^c_\alpha(E') {A^c_\beta
(E')}^*}{E-E'} - i \pi  \sum_{c\,({\rm open})} A^c_\alpha(E)
{A^c_\beta(E)}^*\,.                            \label{6}
\end{equation}
Thus, the effective Hamiltonian (\ref{heff}) can be written as
\begin{equation}
{\cal H}(E)=H+\Delta (E) -\frac{i}{2}\,W (E)\,,
                                               \label{heff}
\end{equation}
which contains an energy-dependent self-energy term due to the
``off-shell" processes of virtual excitation into all, open and
closed at given energy, continuum channels,
\begin{equation}
\langle \alpha |\Delta (E)|\beta\rangle = \sum_{c} \,P \int dE' \,
\frac{A^c_\alpha(E')\, {A^c_\beta(E')}^*}{E-E'},  \label{8}
\end{equation}
and explicitly non-Hermitian and energy dependent term
\begin{equation}
\langle \alpha | W(E)|\beta\rangle = 2\pi  \sum_{c\,({\rm open})}
A^c_\alpha(E)\,{A^c_\beta (E)}^* .                  \label{3}
\end{equation}
The imaginary part $W$ has a factorized form, and originates from
the real ``on-shell" processes through the continuum channels that
are open at given energy. The schematic diagrams representing
various terms in (\ref{heff}) are shown in Fig. \ref{scheme1}.

For our current discussion we will ignore the energy dependence of
the Hamiltonian (\ref{heff}). In general, the energy dependence of
$\Delta$ is smooth and frequently this Hermitian term can be
incorporated into the intrinsic Hamiltonian $H$ as a
renormalization. The energy dependence of the non-Hermitian part
$W$ can be crucial, in particular near channel thresholds, see for
example Wigner's work \cite{wigner48}. In loosely bound systems
this term determines the exact binding energy \cite{continuum}.
Here, however, we are going to discuss the new effects and
especially the phase transition generated by the presence of
non-Hermitian terms in the Hamiltonian, and for this purpose the
amplitudes $A_\alpha^c$ can be taken as energy-independent
parameters, as it would take place in a physical situation with
remote thresholds. We simply assume that the Hamiltonian ${\cal
H}=H-i W /2 $ is known, the anti-Hermitian part has a factorized
structure (\ref{3}), and this will allow us to study the resulting
dynamics.

\section{Two spin-system}
\label{sec:spins}

We start with a very simple example similar to the system
discussed by Dicke \cite{dicke54} in relation to superradiance.
Consider two interacting distinguishable spin-$1/2$ molecules,
$s_1=s_2=s=1/2\,,$ with the spin-spin interaction
\begin{equation}
H^\circ=\alpha \,\vec{s}_1\cdot \vec{s}_2.         \label{10}
\end{equation}
This Hamiltonian is diagonal in the basis of the total spin
$\vec{S}=\vec{s}_1+\vec{s}_2$ and its $z$-projection, and the
energies of the states,
\begin{equation}
E_{S}=\frac{\alpha}{2} \left [S(S+1) - 2 s(s+1) \right ]\,,
                                                  \label{H0}
\end{equation}
are $E_{0}=-3 \alpha /4$ for the singlet, $S=0$, state and
$E_{1}=\alpha /4$ for the triplet, $S=1$, state.

Now let us place the system in the magnetic field that produces
two effects. First, it leads to the Zeeman splitting that can be
described by the additional term in the Hamiltonian
\begin{equation}
H^{\rm B}=\epsilon s_1^z + \epsilon s_2^z = \epsilon S^z.
                                               \label{HB}
\end{equation}
For simplicity we assume that the field acts in the same way on
both molecules, therefore the additional term (\ref{HB}) still
commutes with the operator  $\vec{S}^2.$ Secondly, we assume that
this two-spin system in the magnetic field becomes open; in the
presence of the field the first molecule in its excited polarized
state can dissociate, the phenomenon similar to the Feshbach
resonance \cite{stwalley76}. This means that the molecule in
$s_1^z=1/2$ state decays exponentially with half-life $T_{1/2}$
which can be described by the width $\gamma=\ln 2 /T_{1/2}$. In
terms of the general formalism of Sec.~\ref{sec:heff} here we have
one open decay channel with coupling strength $A$ and
$\gamma=A^2$. In principle the decay rate of a molecule should
depend on the magnitude of the Zeeman splitting and thus on the
field strength. For the interacting system the decay clearly
should depend on total energy. As we stated earlier, we ignore
this dependence in our examples. The situation can be modelled by
an extra non-Hermitian term in the Hamiltonian
\begin{equation}
W=-i \frac{\gamma}{4}(s_1^z+s),
                                               \label{HW}
\end{equation}
so that the first molecule in the state with $s_1^z=1/2$ would
have decay width $\gamma$ and this width would be zero in the
state $s_1^z=-1/2.$ As a result, in the magnetic field the
effective Hamiltonian for the two-spin system becomes
\begin{equation}
{\cal H}= \alpha \,\vec{s}_1\cdot \vec{s}_2 + \epsilon s_1^z +
\epsilon s_2^z-i \frac{\gamma}{4}(s_1^z+s).
                                               \label{HFULL}
\end{equation}

The breakup of the first molecule distorts the symmetry of the
system, $\vec{S}^2$ no longer commutes with ${\cal H}$, and the
singlet and triplet states become mixed. The situation here is
interesting in the sense that the open decay channel introduces an
incompatible symmetry: the interaction Hamiltonian (\ref{H0}) and
the magnetic field part (\ref{HB}) are both diagonal in the basis
of the total spin ($S\, ,S^z\, ,s_1\, ,s_2$), whereas the decay
part, as that of an intrinsic elementary process of the molecule
1, is diagonal in the decoupled spin states of molecules ($s_1\,
,s_1^z\, ; s_2\, ,s_2^z$), the so-called $m$-scheme basis. The
axial symmetry with respect to the magnetic field holds for the
full Hamiltonian (\ref{HFULL}) requiring the conservation of
$S^z.$

\begin{figure}
\begin{center}
\includegraphics[width=6 cm]{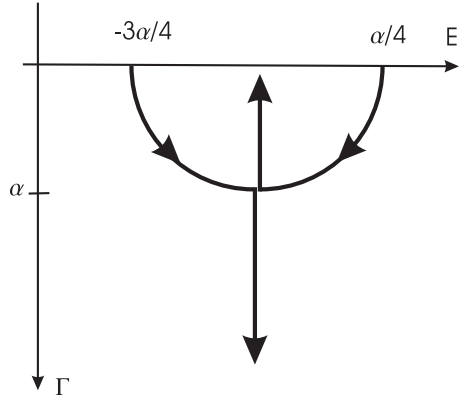}
\end{center}
\caption{\label{scheme2}Evolution of two states ${\cal E}=E-i \Gamma/2$ 
in the complex
plane as a function of the increasing parameter $\gamma.$}
\end{figure}

The full effective Hamiltonian matrix of the $S^z=0$ block in the
$m$-scheme basis takes the form
\begin{equation}
{\cal H}=-\frac{\alpha}{4}+\frac{1}{2} \left (
\begin{array}{cc}
- i \gamma & \alpha \\
\alpha & 0 \\
\end{array}
\right ),
                                          \label{heff2l}
\end{equation}
and the diagonalization of this matrix determines the
quasistationary eigenstates of the system and their complex
energies,
\begin{equation}
{\cal E}_\pm=-\frac{\alpha}{4} \pm \frac{1}{2} \,\sqrt{\alpha^2 -
\left (\frac{\gamma}{2}\right )^2 }\, -i \frac{\gamma}{4}.
                                          \label{2en}
\end{equation}
The trajectories of the complex energies ${\cal E}_\pm$ as functions
of the increasing decay parameter $\gamma$ are shown in Fig.
\ref{scheme2}.

\begin{figure}
\begin{center}
\includegraphics[width=6 cm]{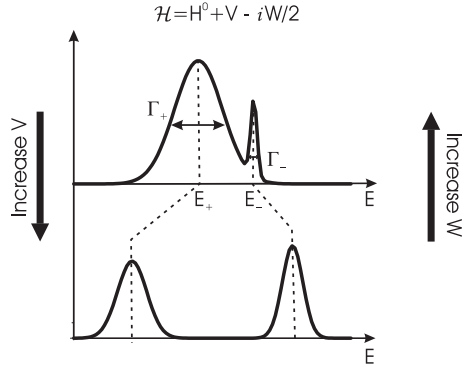}
\end{center}
\caption{\label{two_peaks} Generic behavior of two resonances $E_\pm$ and
their widths $\Gamma_\pm$ as a
result of increase of the Hermitian part of mixing interaction $V$
and in response to increase of non-Hermitian mixing $W.$}
\end{figure}

The dynamics of the two-level effective Hamiltonians have been
extensively studied in the past, see for example
\cite{sokolov94,brentano96,philipp00, brentano02}. In fact the
discussion was initiated by Wigner and von Neumann
\cite{neumann29}, who demonstrated that mixing by an off-diagonal
Hermitian interaction leads to the level repulsion. The behavior
of complex eigenstates as a function of real and complex parts of
the effective Hamiltonian can be summarized as follows. It is well
known that the increase of the Hermitian part of the interaction
leads to repulsion of real energies $E$ and attraction of widths
$\Gamma$ since the decay probability is shared by the two mixed
states. It is less well known that the non-Hermitian part has an
opposite effect, it increases the difference between widths and
moves real parts $E$ closer together
\cite{sokolov94,brentano96,brentano02}; these effects are
schematically demonstrated in Fig. \ref{two_peaks}
\cite{brentano96}.

Our example reveals the generic features, consider  Eq.
(\ref{2en}) and Fig. \ref{scheme2}. In the region of weak decay,
i.e. slow process of molecular disintegration, $0 \le \gamma \le
2\alpha$, the two eigenstates ${\cal E}_+$ and ${\cal E}_-$ have
equal widths, $\Gamma_+=\Gamma_-=\gamma/2$, and the real energies
$E_+$ and $E_-$ are attracted to each other as $\gamma$ grows. At
the critical point $\gamma=2\alpha$ the levels cross and the
``phase transition'' occurs as a sharp change in the behavior of
the two complex energies. The real parts become degenerate,
$E_+=E_-=-\alpha/4,$ while the widths separate,
$\Gamma_{\pm}=\gamma/2 \pm \sqrt{(\gamma/2)^2-\alpha^2}$ so that
in the limit of strong decay, $\gamma\gg \alpha$, we have a single
``superradiant" state with the width $\Gamma_+\rightarrow \gamma$
accumulating the entire available decay probability, and one
long-lived state with $\Gamma_- \rightarrow 0.$ The factorized
structure of matrix $W$ (\ref{3}), that results in zero
eigenvalues in the limit when this part dominates the Hamiltonian,
is a key feature responsible for this behavior. As can be seen
from the derivation of Sec.~\ref{sec:heff}, 
this structure is related to the
analytical structure of the problem and unitarity requirements.
The formation of fast and slow decay modes, which can be viewed as
an ultimate width repulsion, is a typical effect in systems with
overlapping resonances \cite{sokolov88,sokolov89}. In particular,
a detailed discussion for two-level examples can be found in
\cite{sokolov94,philipp00,brentano02,desouter95} and references
therein.

The coupling to decay, as we see, has important observable
consequences for both the intrinsic state of the system and the
external scattering picture. For example, for a singlet ground
state ($\alpha>0$), the magnetic field does not affect the system
in the absence of molecular breakup, because $H_{\rm B}$ is
identically zero in the singlet state. In the presence of the
field, the disintegration of the first molecule reorients the
eigenvector of the system admixing a triplet state. Thus, the system
becomes susceptible to an excitation by the magnetic pulse.

Looking at the situation from a different angle, namely from the
``exterior'', it is clear that the decay of a first molecule alone
is different from the decay properties when the second molecule is
present. In the first paper on superradiation \cite{dicke54} Dicke
argued that coupling to the radiation field can lead to strong
modification of the intrinsic properties of the system. In context
of nuclear reaction theory, a significant contribution to this
subject was made by Wigner and his collaborators who discussed the
nature of resonances in the nucleon-nucleus interactions, sum
rules related to the widths, distribution of the widths among the
states and general features of the scattering cross section
\cite{lane55,teichmann52,teichmann50}, that in context of our
example can be related to the properties of the matrix
(\ref{heff2l}).

\section{A model of an open many-body system}
\label{sec:big}

\begin{figure}
\begin{center}
\includegraphics[width=6 cm]{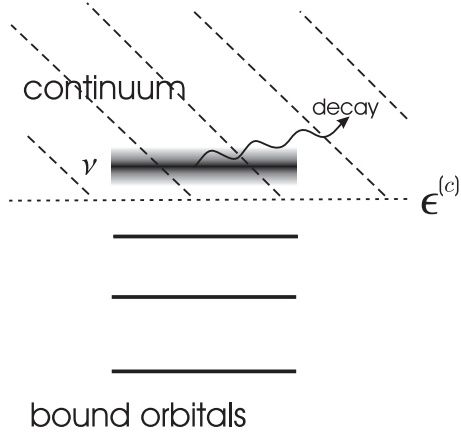}
\end{center}
\caption{\label{scheme3}Schematic drawing of the single-particle
states in the model of Sec.~\ref{sec:big}. The upper state belongs
to the continuum. }
\end{figure}

The transition to superradiance in a many-body model with decay
and random internal dynamics is our second example. We consider an
equidistant set of $\Omega$ fermionic single-particle levels
(``orbitals") and assume that the upper level, that we label as
$\nu$, belongs to the continuum, i.e., in the approximation of
non-interacting particles, the single-particle wave function of
this orbital decreases exponentially with time, while the
remaining orbitals are particle-stable, see Fig. \ref{scheme3}.
This situation again corresponds to a single decay channel where
any many-body state with an occupied single-particle orbital $\nu$
is coupled to the continuum with the corresponding amplitude $A.$
Equivalently the full non-Hermitian effective Hamiltonian ${\cal
H}$ can be constructed from (\ref{hamiltonian}) under assumption
that one of the single-particle energies is complex,
\begin{equation}
e_\nu=\epsilon_\nu-\frac{i}{2}\,\gamma\,,\quad \gamma=A^2\,;
                                                   \label{W}
\end{equation}
it contains single-particle energies with the scale determined by
their spacing $\Delta\epsilon\,,$ the non-Hermitian part $W$, the
scale of which is determined by $\gamma,$ and the Hermitian
mixing $V,$ Eq. (\ref{hamiltonian}). 
We do not impose any symmetry here; the effects of
Kramers degeneracy and rotational invariance have been discussed
in \cite{continuum}. For our example we select the parameters of
the Hermitian residual two-body interaction $V$, eq.
(\ref{hamiltonian}) at random from the Gaussian distribution
centered at zero. The average absolute magnitude of these matrix
elements $\sqrt{\overline{V^2}}$ sets the energy scale for the
model. In what follows we assume $\sqrt{\overline{V^2}}=1$ and
allow $\Delta\epsilon\,$ and $\gamma$ to vary.

\subsection{Evolution of many-body states}

\begin{figure}
\begin{center}
\includegraphics[width=8 cm]{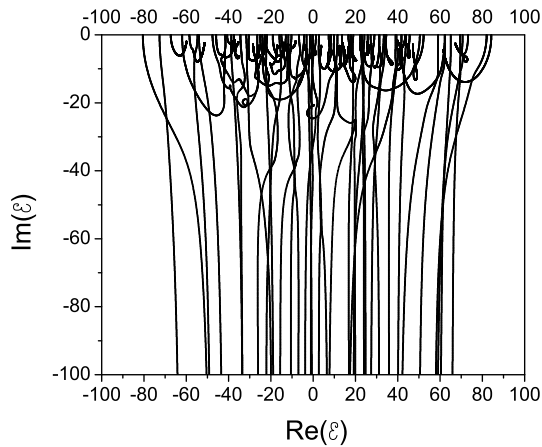}
\end{center}
\caption{Complex plane trajectories of all 70 many-body states in
the system of 4 particles in 8 equidistant single-particle levels
with the spacing of $\Delta\epsilon=0.5$ energy units as a
function of the decay width $\gamma$ of the upper single-particle
state increasing downwards. The two-body interaction matrix
elements are taken randomly from the Gaussian distribution with
zero mean and variance of one energy unit. \label{fig1}}
\end{figure}

A system of $N$ fermions in $\Omega$ single-particle states has
\begin{equation}
{\cal N}=\frac{\Omega!}{N!(\Omega-N)!}       \label{19}
\end{equation}
many-body states. To be specific, we consider $N=4$ particles in
$\Omega=8$ single-particle orbitals. The ${\cal N}=70$ eigenvalues
${\cal E}_{j}=E_{j}-(i/2)\Gamma_{j}$ of the Hamiltonian are in the
lower part of the complex plane, and their motion as a function of
the increasing parameter $\gamma$ is shown in Fig. \ref{fig1};
here $\Delta\epsilon=1$. This figure is a generalization of the
situation in Fig. \ref{scheme2} discussed in the previous section
\ref{sec:spins}. As stated earlier, it is assumed that ${\cal H}$
is independent of energy $E\,$. Therefore the trace conservation
leads to a helpful sum rule for the imaginary part of the
Hamiltonian,
\begin{equation}
\sum_j \Gamma_j(\gamma) = -2 \,{\rm Im}({\rm Tr}\,{\cal H}).
                                             \label{sumgamma}
\end{equation}
This sum rule is a direct analog to the sum rules discussed by Wigner and
collaborators in the context of neutron resonances \cite{lane55}.

Several dynamical limits can be identified. In the ``shell-model'' 
limit of $\gamma=0$ and ${\cal H}=H$, standard diagonalization
results in orthonormal eigenstates $\Psi_{\rm s.m.}$ with real
eigenvalues $E_{\rm s.m.}$. In Fig. \ref{fig1} the values $E_{{\rm
s.m.}}$ define the starting points on the real axis. As soon as
the single-particle width $\gamma\neq 0$, the decay channels open.
The random residual interaction $V$ leads to a chaotic mixing of
configurations, and all many-body states acquire widths. This
excludes special cases when configurations with the occupied
decaying state are blocked, for example by selection rules.

For small $\gamma$, the term $W$ in the Hamiltonian can be treated
perturbatively. In the lowest order this results in
\begin{equation}
{\cal E}=E_{\rm s.m.} - \frac{i}{2}\,\langle \Psi_{\rm s.m}|W|
\Psi_{\rm s.m} \rangle \,.               \label{21}
\end{equation}
Then the decay width of a many-body state is determined by the
occupancy $n_{\nu}(\Psi)=\langle \Psi_{\rm s.m}|a^\dagger_\nu
a_\nu| \Psi_{\rm s.m} \rangle$ of a decaying orbital $\nu$
calculated for a particular shell model state $\Psi_{{\rm s.m.}}$,
\begin{equation}
\Gamma(\Psi)= \langle \Psi_{\rm s.m}|W|
\Psi_{\rm s.m} \rangle = \gamma\, n_\nu(\Psi_{\rm s.m.})\,.
                                              \label{22}
\end{equation}
Only in this limit it is possible to establish the ``natural"
result (\ref{22}) that is widely used in extracting spectroscopic
factors of intrinsic states from various reaction cross sections.

The opposite limit of strong $\gamma$ is less trivial. It is clear
from Fig. 1 that all many-body states are divided into two groups.
In the first group the complex energies have a rapidly increasing
width, whereas the states in the second group become long-lived
with their widths approaching zero. The dominance of the external
coupling $W$ is the origin of this dynamical separation. The
single-particle energies and mixing $V$ are of minor importance in
this limit, and the states can be classified by their relation to
the decay. The matrix $W$ is highly degenerate; because of its
factorized structure, Eq. (\ref{3}), it has only
\begin{equation}
{\cal N}_\Gamma=\frac{(\Omega-1)!}{(N-1)!(\Omega-N)!} \label{23}
\end{equation}
nonzero eigenvalues. The corresponding eigenstates of $W$ are the
configurations that have the decaying upper orbital occupied and
thus acquire in the strong decay limit the maximum width equal to
$\gamma$\, so that the trace (\ref{sumgamma}) of the imaginary
part of the Hamiltonian is exhausted by $\gamma{\cal N}_{\Gamma}.$
In our case, the total dimension of the many-body space is ${\cal
N}=70$, and ${\cal N}_\Gamma=35$. The remaining 35 states
correspond to the empty upper orbital, and correspondingly to the
zero eigenvalues of $W,$ so that decay of these states is
strongly hindered.

Thus, in the limit of strong coupling to the continuum, a
dynamical segregation of short-lived and long-lived (compound)
intrinsic states occurs. This is a generalization of the effect
observed in a simple two-level system of Sec.~\ref{sec:spins}.
This phenomenon was discovered in numerical simulations
\cite{moldauer75,kleinwachter85}, see also
\cite{haake94,persson99}, was studied in molecular and atomic
physics \cite{pavlov88,coleman82,knight90} and observed in
experiments with microwave cavities \cite{persson00}. The fact
that the underlying physics is analogous to the Dicke coherent
state \cite{dicke54} was explained in Refs.
\cite{sokolov88,sokolov89}. The coupling of states via a common
decay channel (width collectivization) plays a role similar to
that of the common radiation field of spontaneously radiating
atoms for the Dicke superradiant coherent mode experimentally
observed in quantum optics \cite{skribanowitz73,gross76}.

A related effect considered in the quantum theory of measurements
has possible ramifications for quantum computers. An external
influence on the system brought by a procedure of measurement, the
Zeno effect \cite{kofman00}, can be modelled similarly, with the
aid of non-Hermitian terms in the Hamiltonian. Recently this topic
attracted a lot of attention due to the appearance of remarkable
experimental data \cite{kofman00,kofman01,fischer01}. In this
context it was demonstrated by Facchi and Pascazio \cite{facchi02}
that an analogous effective subspace separation plays a vital
role. We need to emphasize that the dominating term in the
Hamiltonian, $W,$ because of its degeneracy, only effectively
decouples the subspaces but it does not fully determine the
dynamics. The single-particle energies and residual Hermitian
interactions determine the motion within the intrinsic subspace
that still can be chaotic.

The situation at intermediate values of the decay strength, as
seen from Fig. \ref{fig1}, is complicated. In this region all
three terms of the effective Hamiltonian are important, and their
interplay is responsible for chaotic transitional dynamics. As
$\gamma$ increases, the perturbative approach breaks down.
However, at this point the limit of the full subspace decoupling
is not yet reached. In this regime we are faced with a peculiar
phase crossover, where all components of the dynamics are strongly
mixed and thus both energy centroids $E_{j}$ and widths
$\Gamma_{j}$ become affected in a non-trivial way by the presence
of the non-Hermitian component. A similar interplay between
single-particle energies and the residual pairing interaction is
known to result in chaotic dynamics of the
normal-to-superconducting phase transition in mesoscopic systems
\cite{RP02} which, due to the finite nature of the system, occurs
as an extended crossover.

In his work related to nucleon-nucleus interaction \cite{lane55},
Wigner emphasized the importance of interplay between the state of
intrinsic configuration (determined by Hermitian interaction $V$)
and doorway states (the eigenstates of $W$). He pointed out that
three types of behavior can be distinguished by the distribution
of the widths among many-body states. In the limit of
$V\rightarrow 0$, the ``independent particle model,'' the entire
width is absorbed by the set of states where the decaying
single-particle orbital is fully occupied. This is an analog to
the superradiant regime. In the intermediate regime, due to the
interaction $V$, the width becomes fragmented and spreads to many
states. Finally, in the limit of large $V$ the chaotic intrinsic
motion leads to the approximately uniform distribution of the
widths, the ``uniform model,'' which is analogous here to the
shell-model limit $\gamma\rightarrow 0.$

\subsection{Effective occupation numbers}

We can extend the definition of the single-particle occupancy
$n_{\nu}$ of the decaying orbital to any value of $\gamma$,
instead of the $\gamma\rightarrow 0$ limit discussed previously,
eq. (\ref{22}). The effective occupation number for the
quasistationary state $j$ can be defined as
\begin{equation}
n_{\nu}(j;\gamma) =\frac{\partial \Gamma_{j}(\gamma)}{\partial
\gamma}\,.                                    \label{dynocc}
\end{equation}
As follows from Eq. (\ref{sumgamma}), the sum $\sum_j n_\nu(j;
\gamma)$ is independent of $\gamma\,.$ The numbers (\ref{dynocc}),
properly normalized according to Eq. (\ref{sumgamma}), show how
additional decay amplitudes induced by an infinitesimal increase
of $\gamma$ are distributed among the widths of the eigenstates.
Despite the resemblance to occupation numbers, these $n_\nu$ can
be negative and do not have to be confined to the interval between
0 and 1. There exists also another possibility, namely, to
introduce spectroscopic factors directly as
$\Gamma_j(\gamma)/\gamma\,.$ They are indeed between 0 and 1, but
being simply proportional to $\Gamma$ they do not reveal any new
information related to the dynamics.

\begin{figure}
\begin{center}
\includegraphics[width=8 cm]{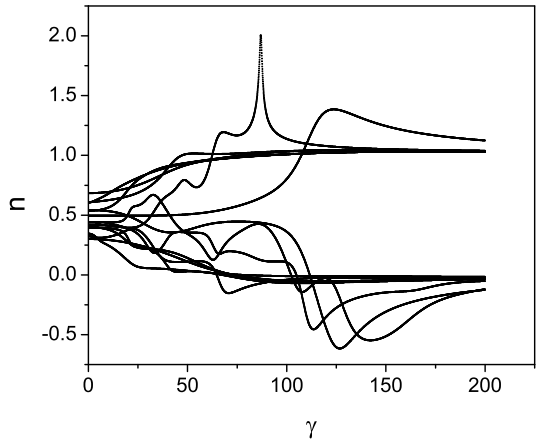}
\end{center}
\caption{The evolution of the effective occupation numbers
$n_\nu$, Eq. (\ref{dynocc}), as a function of $\gamma$ for the 13
lowest (selected at $\gamma=0$) many-body states. \label{fig2}}
\end{figure}

In Fig. \ref{fig2} the evolution of the effective occupation
numbers $n_\nu$ for the lowest 13 states $\Psi_{j}$ is shown. The
evolution starts with occupancies found for the shell model with
residual interaction but no decay and follows a complicated path
to the final segregated values of 0 and 1, corresponding to
quasistable and superradiant states, respectively.  The resulting
crossover transition of the entire spectrum is gradual. It starts
at the point where $\gamma$ exceeds the level spacing on the real
axis. Here the resonances start to overlap. Although there are
states that converge to the limiting value relatively fast, the
majority of states undergo a complicated and peculiar evolution
driven by the competition between the intrinsic and external
interactions. Before reaching the regime of developed segregation,
the evolution may involve peaks of occupation numbers for specific
eigenstates. The segregation occurs at the values of width
covering the macroscopic portion of the many-body spectrum, so
that a balance between the external coupling represented by the
imaginary part and the internal motion given by Hermitian part is
reached. An analogous balance between two-level atoms (radiators)
and radiation field is required for the Dicke superradiance. The
nature of the phase transition to superradiance in the Dicke model
was studied in Refs. \cite{wang73,hioe73}.

The definition of the effective occupation numbers in the form
(\ref{dynocc}) provides a special insight into the dynamics of a
phase transition. This definition is similar in spirit to that of
correlational entropy \cite{sokolov98} that shows how fast the
states of the system readjust as a result of changes in the
environment. Here $\gamma$ plays the role of the external
parameter. For regular dynamics the behavior of this entropy is
smooth. However, any phase transition, which may be hindered in a
mesoscopic system, leads to a rapid restructuring of the state and
thus results in a peak in the correlational entropy
\cite{RP02,cejnar01}. Similarly, in our example, in the
transitional region of $\gamma$, the peaks of $n_\nu$ as a
function of $\gamma$ are clearly seen for most of the states
$\Psi_{j}.$ For some states in Fig. 2 the peaks can be sharp, an
indication that a particular many-body state undergoes a sudden
restructuring. It is worth noting that a practically identical
discussion is appropriate for the two-spin example considered in
Sec.~\ref{sec:spins}. There the quantity $n_\nu$ becomes the
effective probability of finding the molecule 1 in the radiating
state with $s^z=1/2.$ For the two mixed states with $S^z=0$ it can
be easily calculated using the definition (\ref{dynocc}) and the
expression for the complex energy (\ref{2en}). Even in the limited
example of two spins the ``phase transition'' (the point of level
crossing) appears as a special bifurcation point.

\subsection{Global transition to superradiance}

\begin{figure}
\begin{center}
\includegraphics[width=8 cm]{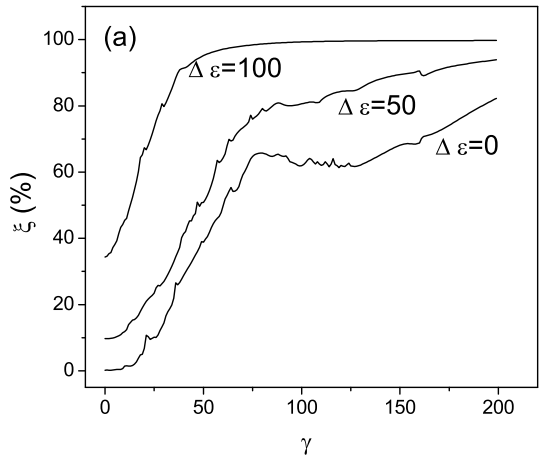}
\end{center}
\caption{Fraction of segregated states as a function of
$\gamma\,.$ The single-particle level spacing is varied from the
degenerate case, $\Delta \epsilon=0$, to $\Delta \epsilon =100$, a
point where the residual interaction $V$ can be completely
ignored. \label{fig3}}
\end{figure}

In Fig. \ref{fig3} we demonstrate the global properties of the
transitional  regime from another perspective by plotting the
segregation fraction $\xi(\gamma)$ defined as a fraction of
many-body states for which the generalized occupancy $n_\nu$
appears close to the limiting values of 0 or 1,
\begin{equation}
\xi(\gamma)=\frac{1}{\cal N}\sum_{j}\left [e^{-n^{2}_\nu(j;\gamma)/(2
\sigma^2)}+e^{-[1-n^{2}_\nu(j;\gamma)]^2/ (2 \sigma^2)} \right ].
                                                  \label{segre}
\end{equation}
Here we choose $\sigma=0.1\,.$ The role of the relative magnitudes
of the single-particle spacing $\Delta \epsilon$ and two-body
matrix elements for the phase transition can be inferred from this
figure. Here a randomly chosen set of two-body matrix elements $V$
with an average magnitude setting the unit of energy is kept
constant, while the single-particle level spacing $\Delta
\epsilon$ is varied. In the limit of infinitely large spacing,
the mixing part of the Hamiltonian, $V$, becomes negligible,
formally resulting in the disappearance of the intermediate
transitional region. Nevertheless, even for $\Delta \epsilon$
fifty times larger than the average $V$, the transitional region
is still well defined.

The transition to superradiance, being gradual for the entire
system, can be very drastic for a particular many-body state. An
interesting analogy can be found between the transition to
superradiance and percolation theory. The set of original
intrinsic states can be considered as a lattice with cites
determined by energies $E_{\rm s.m.}$. Without decay the system is
stationary and rests at a given initial eigenstate. The coupling
to the continuum opens the possibility for the system to hop from
state to state until it gets into a doorway state and emits a
particle. The probability of such hopping $j\rightarrow j'$
according to the perturbation theory is determined by the matrix
element $\langle j|W|j' \rangle $ (this term in our examples is
proportional to $\gamma$) and inversely proportional to the
distance $|{E_{\rm s.m.}}_j-{E_{\rm s.m.}}_{j'}|$. The lattice is
irregular because the cites of the lattice correspond to the
eigenvalues of the Hamiltonian $H$ and the distribution of
distances between the cites is given by Wigner random matrix
theory \cite{wigner55}. 
As a result, some areas can experience phase transitions
and coherently orient toward a doorway state (which is not an
eigenstate) earlier than the rest of the system. This is the
reason for the sharp changes observed in individual states, see
Fig. \ref{fig2} and related discussion. As follows from Fig.
\ref{fig3}, a significant strength of the continuum coupling is
required for the full system to reorient and to create a coherent
superradiant flow into doorway configurations.

\section{Conclusions}
\label{sec:conclusions}

We presented an introduction to interesting phenomena that appear
in complex systems as a result of their coupling to the
environment. We discussed simple examples showing that the strong
coupling significantly reorganizes the motion, dynamically
separating the states by their relation to the external world. We
investigated the properties of a many-body phase transition from
the normal shell model (or Fermi-liquid) behavior of a closed
many-body system to the superradiant regime with the segregation
of rapidly decaying states from long-lived resonances. The chaotic
character of the intrinsic interaction does not prevent the system
from this generic transition. The presence of very long-lived
states, that are almost decoupled from the external world, maybe
important for the problem of quantum computing.

We deliberately related our examples to different branches of
physics and provided many references trying to demonstrate that
these phenomena can be relevant to almost any case when the
coupling of a system to the outside world cannot be neglected.
This topic links together many of the ideas inherited from Wigner.
They include the complexity of many-body states with the characteristic
level repulsion \cite{neumann29}; the theory of quantum damping
and decay \cite{weisskopf30}; the role of symmetry and
incompatibility between symmetries represented here by the
competition between internal symmetry and symmetry imposed by the
decay processes; theory of resonances, resonance sum rules,
fragmentation of the widths in reactions
\cite{lane55,teichmann52,wigner48} and the near-threshold width
behavior \cite{wigner48}; random matrix theory \cite{wigner55}.

\ack
This work was supported by the U. S. Department of Energy, Nuclear
Physics Division, under contract No. W-31-109-ENG-38, and by the
National Science Foundation, grant PHY-0070911. \vskip 1 in


\end{document}